\documentclass[sigconf, screen, nonacm]{acmart} 
\AtBeginDocument{%
  }

\setcopyright{none}

\usepackage{amsmath}
\usepackage{algorithm}
\usepackage{algpseudocode} 
\usepackage{amsfonts}
\usepackage{makecell}

\begin{document}

\title{LLM as HPC Expert: Extending RAG Architecture for HPC Data}

\author{Yusuke Miyashita}
\orcid{0009-0001-3819-2912}
\email{y.miyashita@unsw.edu.au}
\affiliation{%
  \institution{Research Technology Services, University of New South Wales}
  \city{Sydney}
  \state{New South Wales}
  \country{Australia}
}

\author{Patrick Kin Man Tung}
\orcid{0000-0002-2741-3177}
\email{patrick.tung@unsw.edu.au}
\affiliation{%
  \institution{Research Technology Services, University of New South Wales}
  \city{Sydney}
  \state{New South Wales}
  \country{Australia}
}

\author{Johan Barthelemy}
\orcid{0000-0002-7800-5309}
\email{jbarthelemy@nvidia.com}
\affiliation{%
  \institution{NVIDIA}
  \city{Santa Clara}
  \state{California}
  \country{USA}
}

\renewcommand{\shortauthors}{Miyashita et al.}

\begin{abstract}
High-Performance Computing (HPC) is crucial for performing advanced computational tasks, yet their complexity often challenges users, particularly those unfamiliar with HPC-specific commands and workflows. This paper introduces Hypothetical Command Embeddings (HyCE), a novel method that extends Retrieval-Augmented Generation (RAG) by integrating real-time, user-specific HPC data, enhancing accessibility to these systems. HyCE enriches large language models (LLM) with real-time, user-specific HPC information, addressing the limitations of fine-tuned models on such data. We evaluate HyCE using an automated RAG evaluation framework, where the LLM itself creates synthetic questions from the HPC data and serves as a judge, assessing the efficacy of the extended RAG with the evaluation metrics relevant for HPC tasks. Additionally, we tackle essential security concerns, including data privacy and command execution risks, associated with deploying LLMs in HPC environments. This solution provides a scalable and adaptable approach for HPC clusters to leverage LLMs as HPC expert, bridging the gap between users and the complex systems of HPC .
\end{abstract}


\begin{teaserfigure}
  \includegraphics[width=\textwidth]{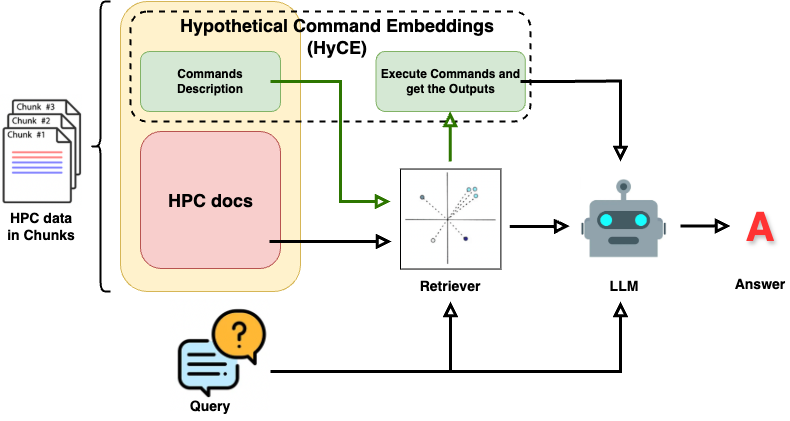}
  \caption{RAG Architecture with HPC Data. In addition to the conventional RAG architecture, it incorporates Hypothetical Command Embeddings (HyCE), which enhances command retrieval to accurately access user-specific HPC data. When a command description is retrieved (indicated by green arrows) based on the query, HyCE executes the command, and the resulting command output is fed into the LLM as context to answer user query}
  \Description{Extended RAG architecture including HyCE, Hypothetical Command Embeddings}
  \label{fig:teaser}
\end{teaserfigure}

\received{20 February 2007}
\received[revised]{12 March 2009}
\received[accepted]{5 June 2009}

\maketitle

\section{Introduction}
HPC is powerful but often inaccessible to a broad range of users due to their inherent complexity. While graphical user interfaces (GUIs) such as Open OnDemand\cite{10.1145/2949550.2949644} have made HPC resources more accessible, users still need to interact with the system through command-line interfaces (CLIs) for resource availability, monitoring queue and software availabilities. This requirement places a steep learning curve on users unfamiliar with shell commands and the intricacies of the HPC environment. Furthermore, relying on HPC experts to answer user queries is not an efficient use of time, as their expertise is a valuable and limited resource. 

Recent advances in large language models (LLMs)\cite{khattab2020colbertefficienteffectivepassage, dubey2024llama3herdmodels} have opened new possibilities for simplifying user interactions with HPC systems. Tools such as ShellGPT\cite{shell_gpt} and AI-Shell\cite{ai_shell} have demonstrated the potential for LLMs to translate human-readable queries into executable shell commands, simplifying the interaction with complex systems. However, these systems lack the capability to integrate cluster-specific documentation or addressing the real-time user information, which is often essential for users to efficiently navigate HPC environments. LLMs directly putting commands in the terminal also poses some security risks. 

Retrieval-Augmented Generation (RAG)\cite{lewis2021retrievalaugmentedgenerationknowledgeintensivenlp} offers a solution by extending LLMs with domain-specific knowledge, dynamically incorporating specific HPC cluster information corresponding to specific HPC organization into the model's context.

Fine-tuned LLMs\cite{Ding_2023, Nichols_2024, chen2024landscapechallengeshpcresearch} have shown potential in aiding parallel programming and detecting race conditions, demonstrating their strength in coding parallel programs. However, this paper focuses on extending RAG to flexibly integrate additional cluster documentation and real-time user information.

We introduce HyCE, an extension of RAG that enriches LLMs with HPC-specific data, allowing them to serve as HPC experts HPC users. To evaluate this approach, we propose a novel framework where the LLM generates synthetic HPC-specific datasets, evaluates itself using defined metrics for HPC workflows, and acts as its own judge to assess performance. Additionally, we address critical security concerns, including safeguarding data privacy and ensuring command execution integrity, to facilitate safe deployment in HPC environments.

This approach not only simplifies access to HPC resources but also lays the foundation for a more interactive and intelligent interface that adapts to the specific needs of each HPC organization. To facilitate this, we open-source the code~\footnote{\url{https://github.com/Yusuke710/llm_rag_eval_hpc}} of our extended RAG, enabling straightforward and secure deployment of LLMs with HPC context in any organization. 

\section{Related Work}
\subsection{LLMs in HPC Environments}
As HPC applications grow in scale and complexity, there are examples of LLMs being adapted to address HPC-specific requirements in programming, code generation\cite{chen2024landscapechallengeshpcresearch,dubey2024llama3herdmodels, Nichols_2024, kadosh2023scopeneedtransformingllms, Chen_2023, kadosh2024monocoderdomainspecificcodelanguage}, and translation from natural language to commands\cite{shell_gpt, ai_shell}. 

HPC-GPT\cite{Ding_2023}, a fine-tuned version of LLaMA\cite{dubey2024llama3herdmodels} on HPC-specific datasets, optimizes tasks like data race detection in OpenMP parallel code, thereby enhancing productivity and accuracy in specialized coding. Similarly, HPC-Coder\cite{Nichols_2024} pushes the boundaries of LLM utility in HPC by generating parallel, performance-optimized code. It assists in annotating code with OpenMP pragmas and forecasting the performance impacts of code modifications, providing expert-level support for complex, high-performance programming. 

To enhance user accessibility to HPC systems, tools like Shell-GPT \cite{shell_gpt} and AI-shell \cite{ai_shell} have been developed to translate natural language queries into executable bash commands. While these tools demonstrate significant potential, academic research on leveraging LLMs to improve user interaction with HPC resources remains limited. This paper addresses this gap by introducing HyCE, a method that integrates HPC commands into a standard RAG architecture. Given the critical importance of secure command execution in HPC environments, this paper also explores the security implications of the proposed approach in a dedicated section.

\subsection{Retrieval-Augmented Generation (RAG) }
RAG\cite{lewis2021retrievalaugmentedgenerationknowledgeintensivenlp} offers another method for incorporating supplementary information into pre-trained LLMs, presenting a flexible alternative to fine-tuning. Unlike fine-tuning, which modifies the model itself, which often is more expensive, RAG dynamically accesses up-to-date or domain-specific information. This dynamic nature makes RAG particularly suited for HPC environments where real-time data, such as user and system information, is essential. For instance, queries like “What GPUs are available to me?” or “What is the status of my program?” necessitate executing HPC commands and interpreting their output within the user’s specific environment. By leveraging RAG, LLMs can respond more accurately, providing context-aware support in HPC contexts. 

Among various RAG techniques, including query translation, indexing, and retrieval\cite{asai2023selfraglearningretrievegenerate, Rackauckas_2024, trivedi2023interleavingretrievalchainofthoughtreasoning}, our HyCE approach took inspiration from Hypothetical Document Embedding (HyDE)\cite{gao2022precisezeroshotdenseretrieval}. HyDE generates a hypothetical answer based on a query, using it to retrieve relevant contexts more accurately. In a similar way, HyCE utilizes hypothetical commands that are embedded into a vector space, which helps in retrieving similar real commands more effectively to enhance the semantic context retrieval.  

Moreover, an automated and continuous evaluation workflow is essential for RAG applications in HPC settings. Recent research has introduced LLMs as evaluators\cite{zheng2023judgingllmasajudgemtbenchchatbot, es2023ragasautomatedevaluationretrieval, adlakha2024evaluatingcorrectnessfaithfulnessinstructionfollowing}, enabling a self-sustaining process in which the LLM generates synthetic data for evaluation and serves as a “judge” to assess RAG performance. Custom evaluation metrics tailored to specific RAG tasks allow LLM evaluators to effectively assess various RAG implementations, including those involving HPC data. 

\begin{figure*}[ht]
    \centering
    \vspace{-1em}
    \includegraphics[width=0.95\linewidth]{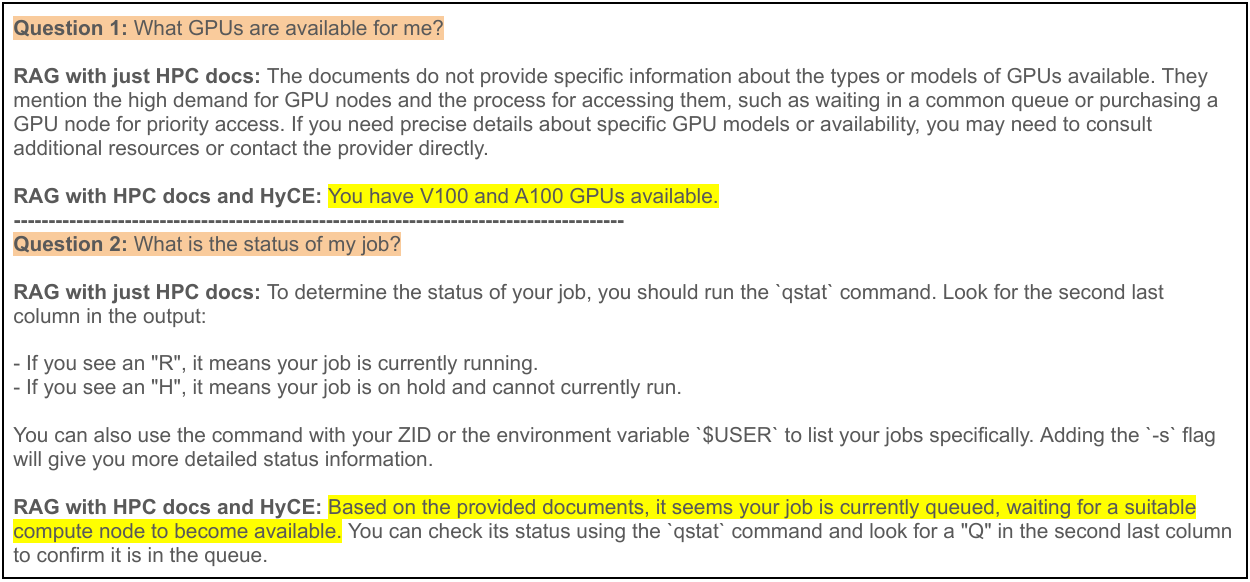}
    \vspace{-0.5em}
    \caption{Example of comparing of two RAG responses. Unlike the ambiguous response provided by RAG with just HPC documents, RAG combined with HyCE can address user specific HPC questions more precisely.}
    \Description{Example of comparing of two RAG responses. Unlike the ambiguous response provided by RAG with just HPC documents, RAG combined with HyCE can address user specific HPC questions more precisely.}
    \label{fig:intro-figure}
    \vspace{-0.5em}
\end{figure*}

\section{RAG with HPC Data }
\subsection{Approach}
In our approach expressed in Figure 1, we extend the standard RAG framework to perform HPC-specific question-answering by integrating data such as cluster documentation and command outputs. In a typical RAG workflow, chunks\cite{günther2024latechunkingcontextualchunk, khattab2020colbertefficienteffectivepassage} are created from these data, embedded with a text embedding model, and used as context. Top K relevant chunks are retrieved with a bi-encoder\cite{devlin2019bertpretrainingdeepbidirectional} and ranked with a cross-encoder\cite{reimers2019sentencebertsentenceembeddingsusing}, then prompt is constructed for LLM to answer. Additionally, we incorporate HyCE into this standard RAG pipeline to accurately include command outputs in the LLM's context. 

\subsection{HPC data}
In this paper, HPC data encompasses to cluster documentations and shell commands.

Cluster documentation serves as a comprehensive guide to navigating an HPC environment. Typically available online, it provides essential details such as the cluster’s purpose, account setup procedures, and access methods. It also includes overviews of hardware, software, and storage resources, along with specific instructions for job submission using schedulers like SLURM or PBS. Furthermore, it outlines best practices for data management and performance optimization while addressing security policies, data protection guidelines, and compliance requirements. Additional resources, such as training materials, support contacts, sample job scripts, tutorials, and FAQs, help users effectively utilize HPC clusters for various tasks.

Shell commands, on the other hand, allow users to retrieve dynamic, real-time information specific to their needs. These commands provide insights into resource availability (e.g., GPUs or CPUs available), job queue status (e.g., running, queued, or held jobs), and software availabilities. Integrating these commands into the RAG framework enables the generation of tailored, context-aware responses, ensuring users receive precise and actionable information for their specific HPC workflows.

\subsection{Hypothetical Command Embeddings (HyCE)}
For handling terminal commands, we employ Hypothetical Command Embeddings (HyCE), an adaptation of the "HyDE" (Hypothetical Document Embeddings) approach. In HyDE, a hypothetical answer to the query is generated to match relevant contexts for retrieval. Similarly, HyCE leverages command descriptions to identify the most appropriate command to execute in response to a query. In this context, a "hypothetical command" refers to a descriptive explanation of what the command does. For example, for the command "nvidia-smi," the description might be "This command checks the GPU model, memory usage, and utilization rate in real time". After retrieving the relevant command, it is executed to generate output, which is then used as context for the LLM to formulate an answer to the query. 

To measure the effectiveness of the HyCE method in retrieving HPC data, we compared the average text similarity between two setups, a cross-encoder which matches the query directly with command and a cross-encoder with HyCE which matches the query with a command description instead. The results indicate the similarity score is consistently higher between the user query and the HyCE-augmented description than between the query and the raw command.

\begin{table*}[ht]
\centering
\caption{Comparison of the average similarity scores between the top-matching query and either the command or command description (HyCE setup). Higher similarity scores indicate greater textual similarity.}
\vspace{-1em}
\label{tab:cross_encoders}
\begin{tabular}{@{}lcc@{}}
\toprule
\textbf{Cross-Encoders}                          & \textbf{\shortstack{Avg Top Sim Score\\Query vs Command Name}} & \textbf{\shortstack{Avg Top Sim Score\\Query vs Command Description}} \\ \midrule
nvidia/llama-3.2-nv-rerankqa-1b-v1            & -2.8283                                                     & \textbf{-2.7880}                                                   \\
cross-encoder/ms-marco-MiniLM-L-12-v2            & -9.0231                                                     & \textbf{-5.1792}                                                   \\
cross-encoder/stsb-roberta-large                 & 0.0906                                                      & \textbf{0.5285}                                                    \\
dangvantuan/CrossEncoder-camembert-large         & 0.2501                                                      & \textbf{0.5718}                                                    \\
yunyu/cross-encoder-stsb-deberta-v3-large        & 0.1472                                                      & \textbf{0.3981}                                                    \\ \bottomrule
\end{tabular}
\end{table*}

Matching the query to a command description proves more accurate than direct matching for several reasons. First, commands are often abbreviated, like "ls" for "list files," which lacks semantic meaning on its own, making direct semantic matching ineffective. Second, commands are syntactically structured and thus incompatible with the semantic search methods typically used in bi-encoder and cross-encoder models. By using descriptive explanations instead, HyCE  makes the command more semantically aligned with the user’s natural language query. 

\section{Automatic RAG evaluation with HPC data}
Using the generated chunks and the extended RAG architecture, we expand the evaluation process to assess RAG's performance with HPC data automatically as described in table \ref{alg:RAG_evaluation}

\begin{algorithm}[ht]
\caption{Automatic RAG Evaluation}
\label{alg:RAG_evaluation}
\raggedright
\textbf{Input:} HPC data chunks $\mathcal{D}$ \\
\textbf{Output:} RAG Evaluation Score

\begin{algorithmic}[0] 
\State $(Q_{hyp}, A_{hyp}) \gets \texttt{LLM\_Synthetic\_Data\_Gen}(\mathcal{D})$ \Comment{Generate synthetic Q\&A pairs}
\State $(Q'_{hyp}, A'_{hyp}) \gets \texttt{LLM\_Filter}(Q_{hyp}, A_{hyp}, \mathcal{D})$ \Comment{Filter pairs}
\State $A_{pred} \gets \texttt{RAG}(Q'_{hyp}, \mathcal{D})$ \Comment{Generate predicted answers}
\State $\texttt{Eval\_Score} \gets \texttt{LLM\_Eval}(A'_{hyp}, A_{pred}, \mathcal{D})$ \Comment{Evaluate answers}
\end{algorithmic}
\textbf{Return} \texttt{Eval\_Score}

\end{algorithm}

\subsection{Question \& Answer Generation and Filtering}
To comprehensively evaluate the RAG system, we utilize chunks of HPC-specific information to generate hypothetical question-answer (Q\&A) pairs. These pairs simulate potential user queries and their corresponding answers based on the provided context, creating a benchmark for testing the system's performance.

The HPC data chunks comprise two key sources: online documentation and shell commands. For Q\&A generation based on command chunks, the commands are executed, and both the command description and output are used in LLM to create the Q\&A pairs. This ensures the generated pairs accurately reflect the practical use cases and specific details of the HPC environment.

In this study, a total of 100 Q\&A pairs were generated: 90 derived from cluster documentation and 10 from shell commands. To ensure the quality and relevance of these pairs, the LLM evaluates them using three criteria aligned with typical HPC usage:

\begin{itemize}
    \item \textbf{Groundedness}: Ensuring the question can be answered unambiguously using the available context.
    \item \textbf{Relevance}: Verifying the question addresses practical and commonly encountered issues in HPC environments.
    \item \textbf{Standalone}: Confirming the question is understandable without requiring additional background information.
\end{itemize}

Each response is evaluated using a binary scoring system (1 for success, 0 for failure), and only Q\&A pairs scoring 1 on all criteria are included in the evaluation set. This rigorous filtering process ensures the Q\&A pairs are both realistic and applicable, providing a robust foundation for assessing the RAG system. Detailed prompts and output formats for this process are provided in the Appendix.

\subsection{Evaluation of RAG via LLM as a Judge}
Following the generation of hypothetical Q\&A pairs, the RAG system generates responses for each question. These responses are evaluated by comparing them to reference answers using defined scoring criteria:

\begin{itemize}
    \item \textbf{Correctness}: Ensuring that the response accurately matches the reference answer.
    \item \textbf{Faithfulness}: Verifying that the response is free from errors, does not hallucinate, and aligns closely with the given context.
\end{itemize}

This evaluation framework assesses both the factual accuracy and the overall reliability of the RAG system’s responses, emphasizing their applicability to HPC users. Unlike the direct scoring approach used for filtering synthetic Q\&A pairs, this stage employs a reference-based evaluation approach\cite{adlakha2024evaluatingcorrectnessfaithfulnessinstructionfollowing}. Each generated answer is scored against the reference answer using binary metrics, assigning a score of 1 for success and 0 for failure for each criterion.

This process provides a structured and rigorous means of evaluating the RAG system, ensuring its responses meet the practical needs of HPC environments. The full prompt can be found in the Appendix.

\section{Experiment and Discussion}
While the extended RAG system is designed for deployment as a chatbot integrated into HPC GUIs like Open OnDemand to enable seamless and intuitive user interactions, this experiment evaluates its functionality in a command-line environment. Specifically, the RAG evaluation was conducted on the terminal of Katana\cite{katana}, an on-premises HPC cluster managed by Research Technology Services at UNSW Sydney. For LLM inference in the RAG, Nvidia NIM microservices and OpenAI were utilized. Details of the RAG hyperparameters and models used in the experiment are provided in the Appendix.

\subsection{Qualitative Analysis of RAG with HPC Data}
We conducted a qualitative analysis to evaluate the impact of incorporating shell command data via HyCE. For instance, when asked, “What GPUs do I have access to?”, an LLM relying solely on cluster documentation provided only generic information about GPU types. In contrast, our RAG setup with HyCE dynamically retrieved user-specific context by executing relevant commands, accurately identifying the user’s available GPUs, such as Nvidia V100 and A100.

This demonstrates how the integration of shell commands enables RAG to generate precise, tailored responses, enhancing the relevance and utility of its outputs for HPC users. Additional examples are shown in Figure 2.

\subsection{Orthogonality of HyCE to Other RAG Improvement Methods}

\begin{table}[h]
\centering
\caption{Automatic RAG evaluation with incremental changes}
\vspace{-1em}
\label{tab:orthogonal performance of HyCE}
\begin{tabular}{@{}lcc@{}}
\toprule
\textbf{Changes}                          & \textbf{Eval Score} & \textbf{$ \Delta $Eval Score} \\ \midrule
RAG baseline                 & 77.67\%                                       & -                                   \\
+ HyCE            & 82.33\%                                      & 4.66\%                                   \\
+ Prompt Engineering(CoT)         & 83\%                                       &0.67\%                                    \\
\makecell[lt]{+ Better Retrieval, \\ Re-Rank models and LLM}        & 86\%                                       & 3\%                                   \\ \bottomrule
\end{tabular}
\end{table}

Our approach to extending RAG with HPC-specific data complements other RAG improvement methods such as prompt engineering using Chain of Thought(CoT)\cite{wei2023chainofthoughtpromptingelicitsreasoning} and better models. The results of automatic evaluation in table \ref{tab:orthogonal performance of HyCE} demonstrate that the incorporation of these components into the pipeline proportionally enhances the performance of our HPC-augmented RAG system. For instance, while HyCE alone improved the baseline RAG by 4.66\%, the combination of CoT and better models effectively leveraged the shell commands' output provided by HyCE, resulting in incremental performance gains.

In addition, our framework is designed for adaptability, enabling users to integrate fine-tuned models optimized for generating parallel code in HPC environments or to adopt open-source models for enhanced security. This flexibility underscores the robustness of our approach across diverse HPC scenarios, allowing users to tailor the system to their unique needs and security requirements.

Furthermore, the RAG architecture can be seamlessly integrated with analytics tools to enhance user support. For example, analyzing user interactions can help identify frequently asked questions, driving iterative improvements to user documentation. These enhancements, in turn, enrich the RAG’s generative capabilities, creating a “spiral of improvement” where both user document and RAG evolve to better serve users.

\subsection{Limitations of the Automatic RAG Evaluation}
Despite its utility, our automatic RAG evaluation method has some limitations. This approach relies on data chunks to create hypothetical questions and answers, which means that it cannot effectively measure performance outside of the provided documentation chunks or commands. Consequently, the system may exhibit hallucinations when responding to queries that fall outside the scope of the user documentation or command output in the real world. Future work could explore evaluation methods that assess RAG’s ability to generalize beyond specific chunks to further enhance robustness of the RAG system. 

\subsection{Other Context Retrieval Metrics}
While context retrieval quality could also be evaluated through metrics such as accuracy, precision, F1 score and AUC, this aspect remains consistent with conventional RAG implementations and thus is not a primary focus in our project. While these metrics may provide additional insights, they fall outside the scope of the HyCE-specific extensions in our RAG approach.

\section{Security Considerations}
\subsection{Data Privacy}
Protecting sensitive HPC information is paramount, especially when interacting with external servers (e.g., OpenAI). Solutions include using enterprise and local models. For enterprise deployments, privacy can be maintained if the provider does not record input-output data and does not train their LLMs with user data, thereby ensuring secure usage. Additionally, future encryption technologies may allow data to be encrypted before being sent externally and processed by the LLM. Alternatively, hosting local models entirely within the HPC environment enhances security by eliminating the need to transmit data externally. However, this approach has challenges, as local model hosting requires substantial compute resources and can reduce available HPC resources for other tasks. Thus, finding a balance between privacy and efficient HPC utilization remains essential. 

\subsection{Command Execution Security}
To maintain security in HPC environments, we implement several layers of safeguards to prevent direct command execution by the LLM:
\begin{itemize}
    \item \textbf{LLM Safety against Prompt Injection:} LLMs are usually pre-trained to counter prompt injection attacks, ensuring that queries with harmful prompts and outputs are detected, keeping the LLM’s responses within safe operational bounds. From the perspective of building RAG, developers can ensure this safety aspect when choosing the LLM model.
    \item \textbf{Predefined Command Retrieval:} In HyCE, RAG is restricted to retrieving only predefined, validated commands, preventing it from executing arbitrary commands autonomously. This reduces the risk of unintended or unsafe system actions.
    \item \textbf{Restricted User Privileges:} The LLM operates under user-level privileges, limiting access to sensitive system functions. This minimizes the potential impact of any unintended commands, restricting the LLM to non-administrative interactions.
    \item \textbf{Containerization:} Deploying the RAG application in a container isolates LLM processes from the core HPC environment, ensuring that any unexpected behavior remains contained and does not impact overall system stability.
\end{itemize}

These layers of security collectively create a secure framework, allowing safe LLM-assisted interactions within HPC environments. 

\section{Conclusion}
In this paper, we introduced HyCE, a novel approach to extending RAG with HPC-specific data, transforming LLMs into effective HPC expert assistants. By incorporating cluster-specific documentation and command outputs, our method enables LLMs to provide contextually accurate, user-tailored responses that address the unique needs of HPC users. Our evaluation framework further strengthens this system, utilizing the LLM as a judge to automate performance assessments and support continuous improvement. This setup, while robust and adaptable, includes layered security safeguards to prevent unauthorized command execution and maintain data privacy, making it suitable for deployment in an HPC environments. 

\section{acknowledgement}
This research was produced in whole or part by UNSW Sydney researchers and is subject to the UNSW
Intellectual property policy. For the purposes of Open Access, the author has applied a Creative Commons
Attribution CC-BY licence to any Author Accepted Manuscript (AAM) version arising from this submission

\bibliographystyle{ACM-Reference-Format}
\bibliography{reference}


\begin{thebibliography}{24}


\ifx \showCODEN    \undefined \def \showCODEN     #1{\unskip}     \fi
\ifx \showDOI      \undefined \def \showDOI       #1{#1}\fi
\ifx \showISBNx    \undefined \def \showISBNx     #1{\unskip}     \fi
\ifx \showISBNxiii \undefined \def \showISBNxiii  #1{\unskip}     \fi
\ifx \showISSN     \undefined \def \showISSN      #1{\unskip}     \fi
\ifx \showLCCN     \undefined \def \showLCCN      #1{\unskip}     \fi
\ifx \shownote     \undefined \def \shownote      #1{#1}          \fi
\ifx \showarticletitle \undefined \def \showarticletitle #1{#1}   \fi
\ifx \showURL      \undefined \def \showURL       {\relax}        \fi
\providecommand\bibfield[2]{#2}
\providecommand\bibinfo[2]{#2}
\providecommand\natexlab[1]{#1}
\providecommand\showeprint[2][]{arXiv:#2}

\bibitem[Adlakha et~al\mbox{.}(2024)]%
        {adlakha2024evaluatingcorrectnessfaithfulnessinstructionfollowing}
\bibfield{author}{\bibinfo{person}{Vaibhav Adlakha}, \bibinfo{person}{Parishad
  BehnamGhader}, \bibinfo{person}{Xing~Han Lu}, \bibinfo{person}{Nicholas
  Meade}, {and} \bibinfo{person}{Siva Reddy}.} \bibinfo{year}{2024}\natexlab{}.
\newblock \bibinfo{title}{Evaluating Correctness and Faithfulness of
  Instruction-Following Models for Question Answering}.
\newblock
\newblock
\showeprint[arxiv]{2307.16877}~[cs.CL]
\urldef\tempurl%
\url{https://arxiv.org/abs/2307.16877}
\showURL{%
\tempurl}


\bibitem[Asai et~al\mbox{.}(2023)]%
        {asai2023selfraglearningretrievegenerate}
\bibfield{author}{\bibinfo{person}{Akari Asai}, \bibinfo{person}{Zeqiu Wu},
  \bibinfo{person}{Yizhong Wang}, \bibinfo{person}{Avirup Sil}, {and}
  \bibinfo{person}{Hannaneh Hajishirzi}.} \bibinfo{year}{2023}\natexlab{}.
\newblock \bibinfo{title}{Self-RAG: Learning to Retrieve, Generate, and
  Critique through Self-Reflection}.
\newblock
\newblock
\showeprint[arxiv]{2310.11511}~[cs.CL]
\urldef\tempurl%
\url{https://arxiv.org/abs/2310.11511}
\showURL{%
\tempurl}


\bibitem[{Builder.io}(2023)]%
        {ai_shell}
\bibfield{author}{\bibinfo{person}{{Builder.io}}.}
  \bibinfo{year}{2023}\natexlab{}.
\newblock \bibinfo{title}{AI Shell}.
\newblock
  \bibinfo{howpublished}{\url{https://github.com/BuilderIO/ai-shell/tree/main}}.
\newblock
\newblock
\shownote{Accessed: 2024-11-24}.


\bibitem[Chen et~al\mbox{.}(2024)]%
        {chen2024landscapechallengeshpcresearch}
\bibfield{author}{\bibinfo{person}{Le Chen}, \bibinfo{person}{Nesreen~K.
  Ahmed}, \bibinfo{person}{Akash Dutta}, \bibinfo{person}{Arijit
  Bhattacharjee}, \bibinfo{person}{Sixing Yu}, \bibinfo{person}{Quazi~Ishtiaque
  Mahmud}, \bibinfo{person}{Waqwoya Abebe}, \bibinfo{person}{Hung Phan},
  \bibinfo{person}{Aishwarya Sarkar}, \bibinfo{person}{Branden Butler},
  \bibinfo{person}{Niranjan Hasabnis}, \bibinfo{person}{Gal Oren},
  \bibinfo{person}{Vy~A. Vo}, \bibinfo{person}{Juan~Pablo Munoz},
  \bibinfo{person}{Theodore~L. Willke}, \bibinfo{person}{Tim Mattson}, {and}
  \bibinfo{person}{Ali Jannesari}.} \bibinfo{year}{2024}\natexlab{}.
\newblock \bibinfo{title}{The Landscape and Challenges of HPC Research and
  LLMs}.
\newblock
\newblock
\showeprint[arxiv]{2402.02018}~[cs.LG]
\urldef\tempurl%
\url{https://arxiv.org/abs/2402.02018}
\showURL{%
\tempurl}


\bibitem[Chen et~al\mbox{.}(2023)]%
        {Chen_2023}
\bibfield{author}{\bibinfo{person}{Le Chen}, \bibinfo{person}{Pei-Hung Lin},
  \bibinfo{person}{Tristan Vanderbruggen}, \bibinfo{person}{Chunhua Liao},
  \bibinfo{person}{Murali Emani}, {and} \bibinfo{person}{Bronis de Supinski}.}
  \bibinfo{year}{2023}\natexlab{}.
\newblock \bibinfo{booktitle}{\emph{LM4HPC: Towards Effective Language Model
  Application in High-Performance Computing}}.
\newblock \bibinfo{publisher}{Springer Nature Switzerland},
  \bibinfo{pages}{18–33}.
\newblock
\showISBNx{9783031407444}
\showISSN{1611-3349}
\urldef\tempurl%
\url{https://doi.org/10.1007/978-3-031-40744-4_2}
\showDOI{\tempurl}


\bibitem[Devlin et~al\mbox{.}(2019)]%
        {devlin2019bertpretrainingdeepbidirectional}
\bibfield{author}{\bibinfo{person}{Jacob Devlin}, \bibinfo{person}{Ming-Wei
  Chang}, \bibinfo{person}{Kenton Lee}, {and} \bibinfo{person}{Kristina
  Toutanova}.} \bibinfo{year}{2019}\natexlab{}.
\newblock \bibinfo{title}{BERT: Pre-training of Deep Bidirectional Transformers
  for Language Understanding}.
\newblock
\newblock
\showeprint[arxiv]{1810.04805}~[cs.CL]
\urldef\tempurl%
\url{https://arxiv.org/abs/1810.04805}
\showURL{%
\tempurl}


\bibitem[Ding et~al\mbox{.}(2023)]%
        {Ding_2023}
\bibfield{author}{\bibinfo{person}{Xianzhong Ding}, \bibinfo{person}{Le Chen},
  \bibinfo{person}{Murali Emani}, \bibinfo{person}{Chunhua Liao},
  \bibinfo{person}{Pei-Hung Lin}, \bibinfo{person}{Tristan Vanderbruggen},
  \bibinfo{person}{Zhen Xie}, \bibinfo{person}{Alberto Cerpa}, {and}
  \bibinfo{person}{Wan Du}.} \bibinfo{year}{2023}\natexlab{}.
\newblock \showarticletitle{HPC-GPT: Integrating Large Language Model for
  High-Performance Computing}. In \bibinfo{booktitle}{\emph{Proceedings of the
  SC ’23 Workshops of The International Conference on High Performance
  Computing, Network, Storage, and Analysis}} \emph{(\bibinfo{series}{SC-W
  2023})}. \bibinfo{publisher}{ACM}, \bibinfo{pages}{951–960}.
\newblock
\urldef\tempurl%
\url{https://doi.org/10.1145/3624062.3624172}
\showDOI{\tempurl}


\bibitem[Es et~al\mbox{.}(2023)]%
        {es2023ragasautomatedevaluationretrieval}
\bibfield{author}{\bibinfo{person}{Shahul Es}, \bibinfo{person}{Jithin James},
  \bibinfo{person}{Luis Espinosa-Anke}, {and} \bibinfo{person}{Steven
  Schockaert}.} \bibinfo{year}{2023}\natexlab{}.
\newblock \bibinfo{title}{RAGAS: Automated Evaluation of Retrieval Augmented
  Generation}.
\newblock
\newblock
\showeprint[arxiv]{2309.15217}~[cs.CL]
\urldef\tempurl%
\url{https://arxiv.org/abs/2309.15217}
\showURL{%
\tempurl}


\bibitem[et~al.(2024)]%
        {dubey2024llama3herdmodels}
\bibfield{author}{\bibinfo{person}{Abhimanyu~Dubey et al.}}
  \bibinfo{year}{2024}\natexlab{}.
\newblock \bibinfo{title}{The Llama 3 Herd of Models}.
\newblock
\newblock
\showeprint[arxiv]{2407.21783}~[cs.AI]
\urldef\tempurl%
\url{https://arxiv.org/abs/2407.21783}
\showURL{%
\tempurl}


\bibitem[Gao et~al\mbox{.}(2022)]%
        {gao2022precisezeroshotdenseretrieval}
\bibfield{author}{\bibinfo{person}{Luyu Gao}, \bibinfo{person}{Xueguang Ma},
  \bibinfo{person}{Jimmy Lin}, {and} \bibinfo{person}{Jamie Callan}.}
  \bibinfo{year}{2022}\natexlab{}.
\newblock \bibinfo{title}{Precise Zero-Shot Dense Retrieval without Relevance
  Labels}.
\newblock
\newblock
\showeprint[arxiv]{2212.10496}~[cs.IR]
\urldef\tempurl%
\url{https://arxiv.org/abs/2212.10496}
\showURL{%
\tempurl}


\bibitem[Günther et~al\mbox{.}(2024)]%
        {günther2024latechunkingcontextualchunk}
\bibfield{author}{\bibinfo{person}{Michael Günther}, \bibinfo{person}{Isabelle
  Mohr}, \bibinfo{person}{Daniel~James Williams}, \bibinfo{person}{Bo Wang},
  {and} \bibinfo{person}{Han Xiao}.} \bibinfo{year}{2024}\natexlab{}.
\newblock \bibinfo{title}{Late Chunking: Contextual Chunk Embeddings Using
  Long-Context Embedding Models}.
\newblock
\newblock
\showeprint[arxiv]{2409.04701}~[cs.CL]
\urldef\tempurl%
\url{https://arxiv.org/abs/2409.04701}
\showURL{%
\tempurl}


\bibitem[Hudak et~al\mbox{.}(2016)]%
        {10.1145/2949550.2949644}
\bibfield{author}{\bibinfo{person}{David~E. Hudak}, \bibinfo{person}{Douglas
  Johnson}, \bibinfo{person}{Jeremy Nicklas}, \bibinfo{person}{Eric Franz},
  \bibinfo{person}{Brian McMichael}, {and} \bibinfo{person}{Basil Gohar}.}
  \bibinfo{year}{2016}\natexlab{}.
\newblock \showarticletitle{Open OnDemand: Transforming Computational Science
  Through Omnidisciplinary Software Cyberinfrastructure}. In
  \bibinfo{booktitle}{\emph{Proceedings of the XSEDE16 Conference on Diversity,
  Big Data, and Science at Scale}} (Miami, USA)
  \emph{(\bibinfo{series}{XSEDE16})}. \bibinfo{publisher}{Association for
  Computing Machinery}, \bibinfo{address}{New York, NY, USA}, Article
  \bibinfo{articleno}{43}, \bibinfo{numpages}{7}~pages.
\newblock
\showISBNx{9781450347556}
\urldef\tempurl%
\url{https://doi.org/10.1145/2949550.2949644}
\showDOI{\tempurl}


\bibitem[Kadosh et~al\mbox{.}(2024)]%
        {kadosh2024monocoderdomainspecificcodelanguage}
\bibfield{author}{\bibinfo{person}{Tal Kadosh}, \bibinfo{person}{Niranjan
  Hasabnis}, \bibinfo{person}{Vy~A. Vo}, \bibinfo{person}{Nadav Schneider},
  \bibinfo{person}{Neva Krien}, \bibinfo{person}{Mihai Capota},
  \bibinfo{person}{Abdul Wasay}, \bibinfo{person}{Nesreen Ahmed},
  \bibinfo{person}{Ted Willke}, \bibinfo{person}{Guy Tamir},
  \bibinfo{person}{Yuval Pinter}, \bibinfo{person}{Timothy Mattson}, {and}
  \bibinfo{person}{Gal Oren}.} \bibinfo{year}{2024}\natexlab{}.
\newblock \bibinfo{title}{MonoCoder: Domain-Specific Code Language Model for
  HPC Codes and Tasks}.
\newblock
\newblock
\showeprint[arxiv]{2312.13322}~[cs.PL]
\urldef\tempurl%
\url{https://arxiv.org/abs/2312.13322}
\showURL{%
\tempurl}


\bibitem[Kadosh et~al\mbox{.}(2023)]%
        {kadosh2023scopeneedtransformingllms}
\bibfield{author}{\bibinfo{person}{Tal Kadosh}, \bibinfo{person}{Niranjan
  Hasabnis}, \bibinfo{person}{Vy~A. Vo}, \bibinfo{person}{Nadav Schneider},
  \bibinfo{person}{Neva Krien}, \bibinfo{person}{Abdul Wasay},
  \bibinfo{person}{Nesreen Ahmed}, \bibinfo{person}{Ted Willke},
  \bibinfo{person}{Guy Tamir}, \bibinfo{person}{Yuval Pinter},
  \bibinfo{person}{Timothy Mattson}, {and} \bibinfo{person}{Gal Oren}.}
  \bibinfo{year}{2023}\natexlab{}.
\newblock \bibinfo{title}{Scope is all you need: Transforming LLMs for HPC
  Code}.
\newblock
\newblock
\showeprint[arxiv]{2308.09440}~[cs.CL]
\urldef\tempurl%
\url{https://arxiv.org/abs/2308.09440}
\showURL{%
\tempurl}


\bibitem[Khattab and Zaharia(2020)]%
        {khattab2020colbertefficienteffectivepassage}
\bibfield{author}{\bibinfo{person}{Omar Khattab} {and} \bibinfo{person}{Matei
  Zaharia}.} \bibinfo{year}{2020}\natexlab{}.
\newblock \bibinfo{title}{ColBERT: Efficient and Effective Passage Search via
  Contextualized Late Interaction over BERT}.
\newblock
\newblock
\showeprint[arxiv]{2004.12832}~[cs.IR]
\urldef\tempurl%
\url{https://arxiv.org/abs/2004.12832}
\showURL{%
\tempurl}


\bibitem[Lewis et~al\mbox{.}(2021)]%
        {lewis2021retrievalaugmentedgenerationknowledgeintensivenlp}
\bibfield{author}{\bibinfo{person}{Patrick Lewis}, \bibinfo{person}{Ethan
  Perez}, \bibinfo{person}{Aleksandra Piktus}, \bibinfo{person}{Fabio Petroni},
  \bibinfo{person}{Vladimir Karpukhin}, \bibinfo{person}{Naman Goyal},
  \bibinfo{person}{Heinrich Küttler}, \bibinfo{person}{Mike Lewis},
  \bibinfo{person}{Wen tau Yih}, \bibinfo{person}{Tim Rocktäschel},
  \bibinfo{person}{Sebastian Riedel}, {and} \bibinfo{person}{Douwe Kiela}.}
  \bibinfo{year}{2021}\natexlab{}.
\newblock \bibinfo{title}{Retrieval-Augmented Generation for
  Knowledge-Intensive NLP Tasks}.
\newblock
\newblock
\showeprint[arxiv]{2005.11401}~[cs.CL]
\urldef\tempurl%
\url{https://arxiv.org/abs/2005.11401}
\showURL{%
\tempurl}


\bibitem[Nichols et~al\mbox{.}(2024)]%
        {Nichols_2024}
\bibfield{author}{\bibinfo{person}{Daniel Nichols}, \bibinfo{person}{Aniruddha
  Marathe}, \bibinfo{person}{Harshitha Menon}, \bibinfo{person}{Todd Gamblin},
  {and} \bibinfo{person}{Abhinav Bhatele}.} \bibinfo{year}{2024}\natexlab{}.
\newblock \showarticletitle{HPC-Coder: Modeling Parallel Programs using Large
  Language Models}. In \bibinfo{booktitle}{\emph{ISC High Performance 2024
  Research Paper Proceedings (39th International Conference)}}.
  \bibinfo{publisher}{IEEE}, \bibinfo{pages}{1–12}.
\newblock
\urldef\tempurl%
\url{https://doi.org/10.23919/isc.2024.10528929}
\showDOI{\tempurl}


\bibitem[Rackauckas(2024)]%
        {Rackauckas_2024}
\bibfield{author}{\bibinfo{person}{Zackary Rackauckas}.}
  \bibinfo{year}{2024}\natexlab{}.
\newblock \showarticletitle{RAG-Fusion: A New Take on Retrieval Augmented
  Generation}.
\newblock \bibinfo{journal}{\emph{International Journal on Natural Language
  Computing}} \bibinfo{volume}{13}, \bibinfo{number}{1} (\bibinfo{date}{Feb.}
  \bibinfo{year}{2024}), \bibinfo{pages}{37–47}.
\newblock
\showISSN{2319-4111}
\urldef\tempurl%
\url{https://doi.org/10.5121/ijnlc.2024.13103}
\showDOI{\tempurl}


\bibitem[Reimers and Gurevych(2019)]%
        {reimers2019sentencebertsentenceembeddingsusing}
\bibfield{author}{\bibinfo{person}{Nils Reimers} {and} \bibinfo{person}{Iryna
  Gurevych}.} \bibinfo{year}{2019}\natexlab{}.
\newblock \bibinfo{title}{Sentence-BERT: Sentence Embeddings using Siamese
  BERT-Networks}.
\newblock
\newblock
\showeprint[arxiv]{1908.10084}~[cs.CL]
\urldef\tempurl%
\url{https://arxiv.org/abs/1908.10084}
\showURL{%
\tempurl}


\bibitem[{TheR1D}(2024)]%
        {shell_gpt}
\bibfield{author}{\bibinfo{person}{{TheR1D}}.} \bibinfo{year}{2024}\natexlab{}.
\newblock \bibinfo{title}{Shell GPT}.
\newblock \bibinfo{howpublished}{\url{https://github.com/TheR1D/shell_gpt}}.
\newblock
\newblock
\shownote{Accessed: 2024-11-24}.


\bibitem[Trivedi et~al\mbox{.}(2023)]%
        {trivedi2023interleavingretrievalchainofthoughtreasoning}
\bibfield{author}{\bibinfo{person}{Harsh Trivedi}, \bibinfo{person}{Niranjan
  Balasubramanian}, \bibinfo{person}{Tushar Khot}, {and}
  \bibinfo{person}{Ashish Sabharwal}.} \bibinfo{year}{2023}\natexlab{}.
\newblock \bibinfo{title}{Interleaving Retrieval with Chain-of-Thought
  Reasoning for Knowledge-Intensive Multi-Step Questions}.
\newblock
\newblock
\showeprint[arxiv]{2212.10509}~[cs.CL]
\urldef\tempurl%
\url{https://arxiv.org/abs/2212.10509}
\showURL{%
\tempurl}


\bibitem[{University of New South Wales}(2014)]%
        {katana}
\bibfield{author}{\bibinfo{person}{{University of New South Wales}}.}
  \bibinfo{year}{2014}\natexlab{}.
\newblock \bibinfo{title}{Katana}.
\newblock \bibinfo{howpublished}{\url{https://doi.org/10.26190/669x-a286}}.
\newblock
\newblock
\shownote{Accessed: 2024-11-24}.


\bibitem[Wei et~al\mbox{.}(2023)]%
        {wei2023chainofthoughtpromptingelicitsreasoning}
\bibfield{author}{\bibinfo{person}{Jason Wei}, \bibinfo{person}{Xuezhi Wang},
  \bibinfo{person}{Dale Schuurmans}, \bibinfo{person}{Maarten Bosma},
  \bibinfo{person}{Brian Ichter}, \bibinfo{person}{Fei Xia},
  \bibinfo{person}{Ed Chi}, \bibinfo{person}{Quoc Le}, {and}
  \bibinfo{person}{Denny Zhou}.} \bibinfo{year}{2023}\natexlab{}.
\newblock \bibinfo{title}{Chain-of-Thought Prompting Elicits Reasoning in Large
  Language Models}.
\newblock
\newblock
\showeprint[arxiv]{2201.11903}~[cs.CL]
\urldef\tempurl%
\url{https://arxiv.org/abs/2201.11903}
\showURL{%
\tempurl}


\bibitem[Zheng et~al\mbox{.}(2023)]%
        {zheng2023judgingllmasajudgemtbenchchatbot}
\bibfield{author}{\bibinfo{person}{Lianmin Zheng}, \bibinfo{person}{Wei-Lin
  Chiang}, \bibinfo{person}{Ying Sheng}, \bibinfo{person}{Siyuan Zhuang},
  \bibinfo{person}{Zhanghao Wu}, \bibinfo{person}{Yonghao Zhuang},
  \bibinfo{person}{Zi Lin}, \bibinfo{person}{Zhuohan Li},
  \bibinfo{person}{Dacheng Li}, \bibinfo{person}{Eric~P. Xing},
  \bibinfo{person}{Hao Zhang}, \bibinfo{person}{Joseph~E. Gonzalez}, {and}
  \bibinfo{person}{Ion Stoica}.} \bibinfo{year}{2023}\natexlab{}.
\newblock \bibinfo{title}{Judging LLM-as-a-Judge with MT-Bench and Chatbot
  Arena}.
\newblock
\newblock
\showeprint[arxiv]{2306.05685}~[cs.CL]
\urldef\tempurl%
\url{https://arxiv.org/abs/2306.05685}
\showURL{%
\tempurl}


\end{thebibliography}

\appendix
\section{RAG Hyperparameters and Models used}
\begin{table}[ht]
\centering
\label{tab:rag_hyperparameters_models}
\begin{tabular}{@{}p{0.3\linewidth}p{0.65\linewidth}@{}}
\toprule
\multicolumn{2}{c}{\textbf{Hyperparameters}} \\ \midrule
\textbf{Parameter}           & \textbf{Value}                 \\ \midrule
Chunk Size                   & Determined by LLM             \\
Top K Retrieval              & 20                           \\
Top K Re-rank                & 5                            \\
Max Context Size             & 128k tokens (limit of LLM)    \\
LLM Temperature                  & 0                          \\
Max Output Tokens            & 4096                          \\ \midrule
\multicolumn{2}{c}{\textbf{Models}} \\ \midrule
\textbf{Category}            & \textbf{Model Name(increasing order of performance in the RAG)}           \\ \midrule
Retrieval Model             & nvidia/llama-3.2-nv-embedqa-1b-v1 \\
& sentence-transformers/multi-qa-MiniLM-L6-cos-v1 \\
Re-rank Model               & nvidia/llama-3.2-nv-rerankqa-1b-v1     \\
& cross-encoder/ms-marco-MiniLM-L-12-v2     \\
LLM                          & meta/llama-3.1-405b-instruct       \\
                             & gpt-4o-2024-08-06                        \\ \bottomrule
\end{tabular}
\end{table}

\section{LLM Prompt Templates}
We list the prompt templates for LLM Q\&A generation, Q\&A filtering, LLM answer, and LLM judge. Please refer to our GitHub repository~\footnote{\url{https://github.com/Yusuke710/llm_rag_eval_hpc}} for full details.

\begin{figure*}
\centering
\includegraphics[width=\linewidth]{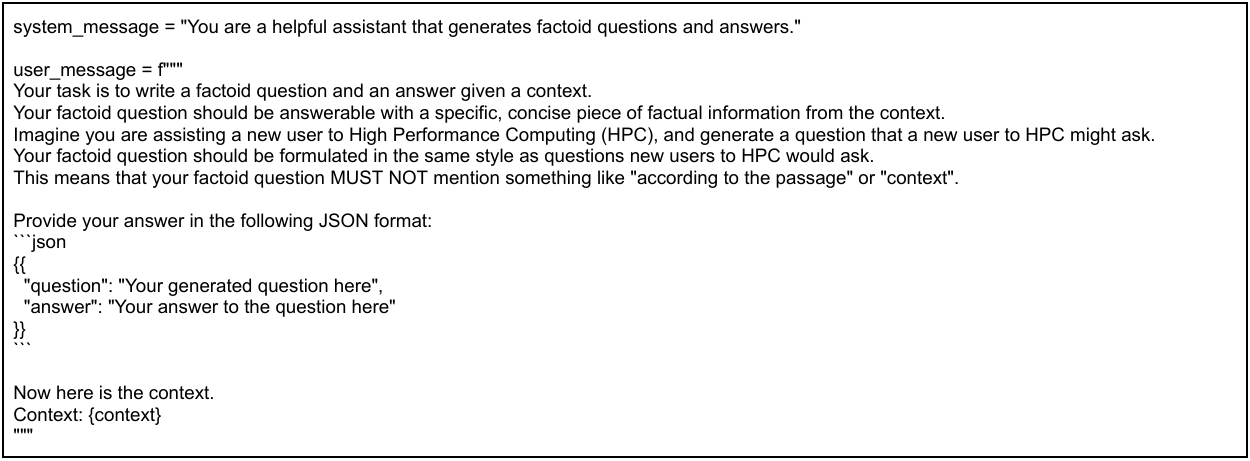}
\caption{The prompts for RAG Q\&A generation}
\Description{The prompts for RAG Q\&A generation}
\label{fig:qa_generation_prompt}
\end{figure*}

\begin{figure*}
\centering
\includegraphics[width=\linewidth]{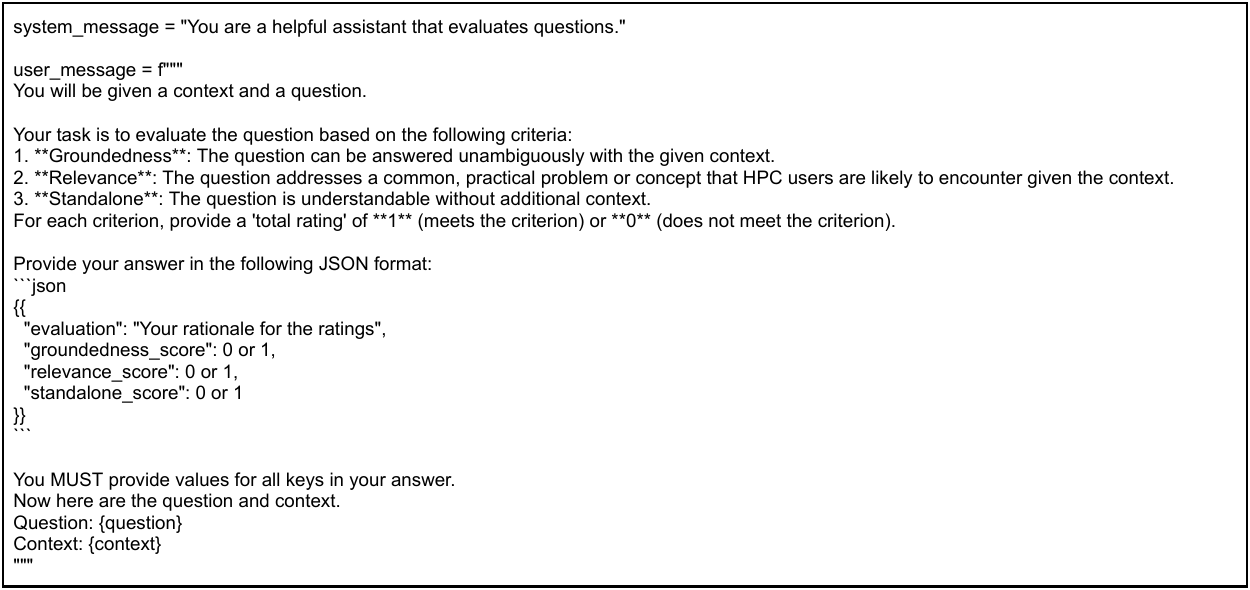}
\caption{The prompts for RAG Q\&A filtering}
\Description{The prompts for RAG Q\&A filtering}
\label{fig:qa_filter_prompt}
\end{figure*}

\begin{figure*}
\centering
\includegraphics[width=\linewidth]{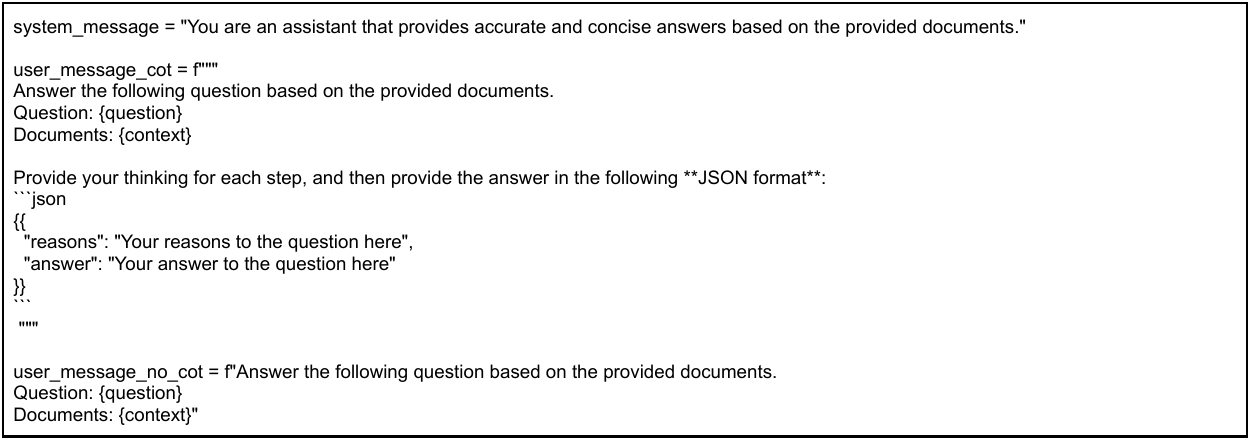}
\caption{The prompts for RAG answering}
\Description{The prompts for RAG answering}
\label{fig:answer_prompt}
\end{figure*}

\begin{figure*}
\centering
\includegraphics[width=\linewidth]{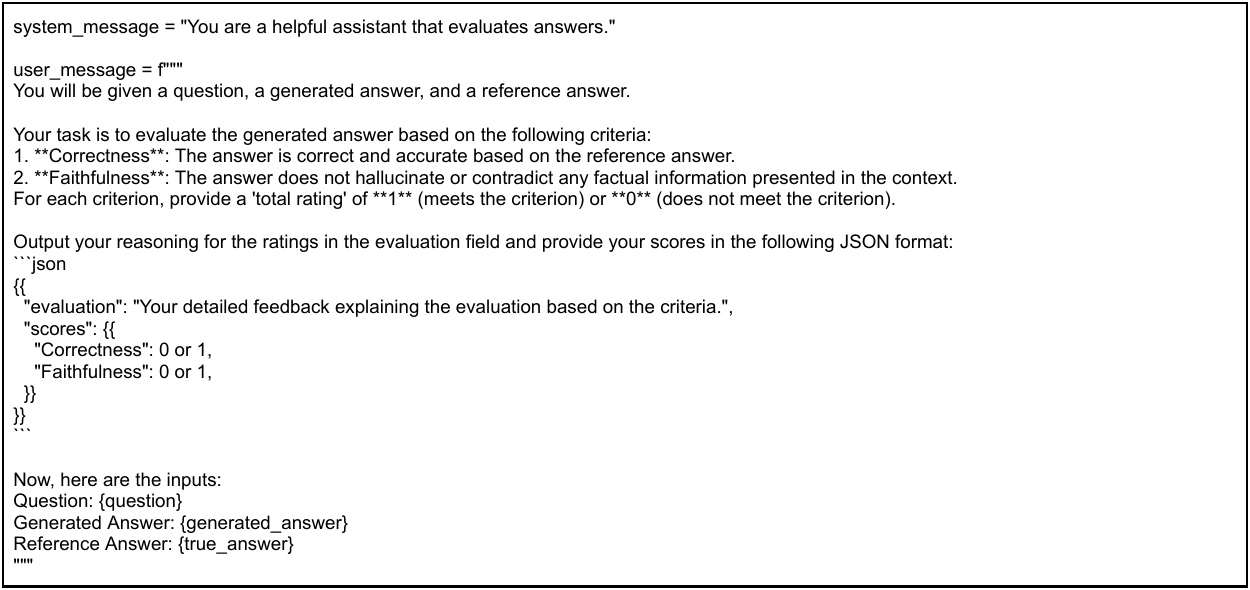}
\caption{The prompts for RAG evaluation}
\Description{The prompts for RAG evaluation}
\label{fig:eval_prompt}
\end{figure*}

\end{document}